\documentclass[12pt]{article}
\usepackage{amssymb,graphicx}
\usepackage{epstopdf}
\usepackage{amsmath,amsfonts}
\usepackage{epsfig}
\usepackage[titletoc,title]{appendix}
\usepackage{xcolor}
\usepackage{amsmath}
\usepackage{amsfonts}
\usepackage{amssymb}

\def\d{\partial}

\newcommand{\be}{\begin{equation}}
\newcommand{\ee}{\end{equation}}
\newcommand{\bea}{\begin{eqnarray}}
\newcommand{\eea}{\end{eqnarray}}
\newcommand{\bg}{\begin{gather}}
\newcommand{\eg}{\end{gather}}
\newcommand{\bseq}{\begin{subequations}}
	\newcommand{\eseq}{\end{subequations}}

\usepackage[english]{babel}
\usepackage{amsmath,amssymb} 
\usepackage{ mathrsfs }
\newcommand{\RNumb}[1]{\uppercase\expandafter{\romannumeral #1\relax}}
\numberwithin{equation}{section}

\begin{document}
	\begin{flushright}
		INR-TH-2020-022
	\end{flushright}
	\vspace{10pt}
	\begin{center}
	  {\LARGE \bf Power-law Genesis: strong coupling
            and  galileon-like vector fields.} \\
		\vspace{20pt}
		%\medskip
	P. K. Petrov$^{a,b}$\\
	\vspace{15pt}
	$^a$\textit{
Department of Particle Physics and Cosmology, Faculty of Physics,
M. V. Lomonosov Moscow State University, Vorobyovy Gory, 1-2, Moscow,
119991, Russia
	}\\
	\vspace{5pt}
	$^b$\textit{
		Institute for Nuclear Research of
		the Russian Academy of Sciences,\\  60th October Anniversary
		Prospect, 7a, 117312 Moscow, Russia}\\
	\vspace{5pt}
	\end{center}
	\vspace{5pt}
	
	\begin{abstract}

                   A simple way to construct models with
          early cosmological Genesis epoch
          is to employ bosonic fields whose Lagrangians
                  transform homogeneously under scaling transformation.
                  We show that                
         in these theories,
         for a range of parameters defining
         the Lagrangian,
         there exists a homogeneous power-law solution
         in flat space-time, whose energy density
         vanishes, while pressure is negative (power-law Genesis).
         We find the condition
         for the legitimacy of the classical field theory description
         of such a situation.
         We note that this condition does not hold for our earlier
         Genesis model
         with vector field. We   construct another model with vector
         field and power-law background solution in flat space-time,
         which is legitimately treated within classical field theory,
         violates the NEC and is stable.
         Upon turning on gravity, this model describes the early Genesis stage.
\end{abstract}

                  \section{Introduction and summary.}
                    
                  Genesis \cite{1} is a cosmological scenario without
                  initial singularity.
                  In this scenario, the Universe starts its
                  expansion from flat space-time
                  and zero energy density
                  at large negative times.
                  As the Universe evolves, the energy density
                  and the Hubble rate        
                  grow, and eventually reach large values.                  
                  If gravity is described by General Relativity, then
                  this regime requires the domination of exotic matter
                  which violates the Null Energy Condition,
                  NEC (for a review see \cite{2}). Later on,
                  the energy density of exotic matter has to be
                  converted into the energy density of usual matter,
                  and the conventional
                  cosmological evolution starts.  As 
                  shown in \cite{3}, the violation of the NEC in
                  a healthy way is possible in the context of the
                  scalar Galileon theories \cite{4}. By now, numerous ways to
                  implement the Genesis idea have been proposed, mostly in the
                  context of theories involving scalars (see
                  Ref.~\cite{1,5} for an incomplete list and Ref.~\cite{Khalat}
                  for topical review), but also in
                  vector field models~\cite{6}.

                  A straighforward way to construct a model of the
                  early Genesis epoch
                  is to make use of a Lagrangian which,
                  in the absence of gravity,
                  transforms homogeneously
                  under 
                  scaling transformation:
                  $\mathscr{L} \Rightarrow \lambda^{N} \mathscr{L}$ when
                  $\pi_{\alpha}(x^{\nu}) \Rightarrow
                  \lambda^{s} \pi_{\alpha}(\lambda x^{\nu})$, where
                  $\pi_\alpha$ denote the non-gravitational fields 
                  in the model, and $N$ and $s$ are constant parameters.
                  Then, quite generally, the model,
                  still
                  in the absence of gravity,
                  has a
                  spatially homogeneous solution $\pi \propto |t|^{-s}$
                  as $t\to -\infty$
                  (power-law Genesis,
                  see Sec.~\ref{sec:strongcoupling}), for which the
                  energy density vanishes while pressure is negative.
                  This precisely means the violation of the NEC. When gravity
                  (described by GR) is
                  turned on, energy density no longer stays equal to zero;
                  instead, it increases as required for the early Genesis
                  stage. This mechanism has been invented in Ref.~\cite{1}
                  (with $N=4$ and $s=1$)
                  and then utilized in other contexts (see Ref.~\cite{2} for
                  a review),
                  including models with vector fields~\cite{6}.

                  However, within this class
                  of models, the coefficients
                  in the quadratic Lagrangian for perturbations
                  about the classical solution often tend to zero as
                  $t\rightarrow-\infty$, which
                  implies that
                  the strong coupling energy scale also tends to zero.
                  In such a situation, the classical treatment may become
                  problematic, cf. Refs.~\cite{1,7}.
                  To figure out whether or not this is the case,
                  one should study both qadratic and interaction terms
                  in the Lagrangian for perturbations and find the behavior
                  of the strong coupling scale $\Lambda$ as $t\to - \infty$:
                   \[
                   \Lambda (t) \propto |t|^{-\sigma} \; .
                   \]
                   This scale
                   should be compared with
                  the classical energy scale
                  $E_{cl}$, which is merely the evolution rate, and
                  in the power-law Genesis case is given by
                   \[
                   E_{cl} (t) \propto |t|^{-1} \; .
                   \]
                  The classical treatment is legitimate provided that
                   $E_{cl} \ll \Lambda$, which means
                   \be
                   \sigma \leq 1
                   \label{apr14-20-1}
                   \ee
                   (the case $\sigma=1$ is subtle: the
                   relation  $E_{cl} \ll \Lambda$
                   may be valid in a restricted region of parameter space).

                   In this note we address this strong coupling issue
                   in the context of the power-like models described above.
                   This is done in Sec.~\ref{sec:strongcoupling},
                   where we show that the requirement \eqref{apr14-20-1}
                   is equivalent to
                   \[
                   N \leq 4 \; .
                   \]
                   We note that this property does {\it not} hold for
                   Genesis with vector field proposed in
                   Ref.~\cite{6}.
                   Therefore, in Sec.~\ref{sec:vectorfield} we construct
                   another model with vector field and power-law background
                   solution
                   that obeys \eqref{apr14-20-1};
                   we determine the range of parameters
                   in which the background is stable and
                   violates the NEC in Minkowski space. For completeness, we
also turn on gravity (in the form of GR) and
describe the evolution of the scale factor at the early
Genesis stage.

 \vspace{0.3 cm}

\section{Strong coupling scale.}
\label{sec:strongcoupling}

As outlined in Introduction, we
consider the Lagrangian for $M$ bosonic fields
$\pi_{\alpha},\;\alpha=1,2,...,M$, in 4d Minkowski space.
Index $\alpha$ may either enumerate
the fields (say, if $\pi_\alpha$ are scalars)
or be Lorentz index, or both. The Lagrangian is assumed to
transform homogeneously
under scaling transformation
$$
x^{\nu} \Rightarrow \lambda x^{\nu}\; , \;\;\;\;\;
\pi_{\alpha}(x^{\nu}) \Rightarrow \lambda^{s} \pi_{\alpha}(\lambda x^{\nu}),
\; \, \;\;\;\;\;\;s\neq 0.
$$
Namely,
\be
\mathscr{L} \Rightarrow \lambda^{N} \mathscr{L}.
\label{1.1}
\ee
Importantly, we assume that equations of motion
are second order in derivatives, even though the Lagrangian
may involve second derivatives of the fields.
This is the case in generalized Galileon theories~\cite{2,4,8} as well
as in theories with Galileon-like vector fields~\cite{6}.

We consider for definiteness the Lagrangians which are
linear combinations of the
monomials involving $n$ fields without derivatives,
$m$ first derivatives and $l$ second derivatives of the fields
(the argument goes through if one allows 
also for inverse powers of the fields):
\be
(\pi_{\alpha_{1}}...\pi_{\alpha_{n}})\cdot 
(\partial \pi_{\gamma_{1}}...\partial \pi_{\gamma_{m}}) 
\cdot (\partial^2 \pi_{\omega_{1}}...\partial^2 \pi_{\omega_{l}})
\sim [\pi]^n \cdot [\d \pi]^m \cdot [\d^2 \pi]^l\; .
\label{mar30-20-1}
\ee
Here
$$
ns+m(s+1)+l(s+2)=N,
$$
so that 
 the transformaton law
\eqref{1.1} holds.
%It is straightforward to see that
 For a range of parameters defining the Lagrangian,
there exists a homogeneous power-law solution
\be
\pi^{{(0)}}_{\alpha}=\beta_{\alpha} |t|^{-s}, 
\label{mar30-20-2}
\ee
with constant $\beta_\alpha$. Indeed, the term \eqref{mar30-20-1}
gives a contribution to equation of motion with total number of
fields equal to $(n+m+l-1)$ and total number of derivatives $(m+2l)$.
Therefore, with the Ansatz \eqref{mar30-20-2}, each of the $M$
equations of motion is proportional to $|t|^{-N+s}$ with the proportionality
coefficient being a polynomial in $\beta_\alpha$. In other words,
equations of motion make a system of
$M$ algebraic equations for $M$ coefficients
$\beta_\alpha$, which has a solution for a range of parameters
entering the Lagrangian \footnote{
    Unless there is some symmetry that
    relates coefficients of different monomials \eqref{mar30-20-1}
    in such a way that this algebraic system does not have a real solution.}.

Let us now consider perturbations about the background
\eqref{mar30-20-2}, $\pi_\alpha = \pi_\alpha^{(0)} +\delta \pi$.
Our purpose is to determine the time-dependence
of the lowest
strong coupling scale in the limit $t\to -\infty$.
We begin with quadratic Lagrangian for perturbations.
Since we assume that there are no third and higher derivatives
of $\delta \pi_\alpha$ in the equations of motion,
there are no terms with second and higher derivatives in the
quadratic Lagrangian. So,
the relevant terms
are,
schematically, $(\d \, \delta \pi)^2$. The monomial \eqref{mar30-20-1}
in the original Lagrangian contributes to the terms
 $(\d \, \delta \pi)^2$ in the quadratic Lagrangian with 
coefficients involving $(n+m+l-2)$ background fields $\pi^{(0)}$ and
$(m+2l-2)$ derivatives acting on them.
Hence, the structure of the quadratic
Lagrangian is
   $$
   {\cal L}^{(2)} \supset |t|^{-N + 2s +2} (\d \, \delta \pi)^2
\; .
   $$
   This implies that canonically normalized fields are
   \be
   \xi_\alpha \propto |t|^{-N/2+s+1} \delta \pi_\alpha \; .
\label{mar30-20-10}
   \ee
Their mass dimension, by definition, equals 1.

 We now turn to the interactions between perturbations
  $\delta \pi$. The term \eqref{mar30-20-1} induces interactions of
  the following form:
  $$
     [\pi^{{(0)}}]^{n-a} \cdot  [\d \pi^{{(0)}}]^{m-b} \cdot [\d^2 \pi^{{(0)}}]^{l-c}
     \times    [\delta \, \pi]^{a}
     \cdot  [\d \, \delta \, \pi]^{b}  \cdot [\d^2 \, \delta\, \pi ]^{c},
     $$
     where
     \be
               a+b+c \geq 3.
\label{mar30-20-11}
     \ee
     We make use of \eqref{mar30-20-2} and \eqref{mar30-20-10}
     and find that in terms of canonically normalized field, this
     interaction Lagrangian is proportional to
     $$
     |t|^{\frac{N}{2}(a+b+c-2) +c -a} \times    [\xi]^{a}
     \cdot  [\d \xi]^{b}  \cdot [\d^2 \xi]^{c}.
     $$
     On dimensional grounds, the coefficient of  $[\xi]^{a}
       \cdot  [\d \xi]^{b}  \cdot [\d^2 \xi]^{c}$ in the Lagrangian
       is $E_s^{-(a+2b+3c -4)}$, where $E_s$ is the (naive) strong
       interaction scale (we consider the case
$a+2b+3c -4 >0$, otherwise no constraint is obtained). Thus,
       $$
       E_s \propto  |t|^{-\frac{\frac{N}{2}(a+b+c-2) +c -a}{a+2b+3c-4}}.
       $$
       We require that this scale is higher than the classical energy scale
       $t^{-1}$ for $|t| \to \infty$ and get
       $$
       \frac{\frac{N}{2}(a+b+c-2) +c -a}{a+2b+3c-4} < 1,
       $$
       or
       $$
        (N-4)(a+b+c-2)<0.
       $$
       We recall \eqref{mar30-20-11}      and obtain finally
       $$
       N \leq 4,
       $$
       where we include the case $N=4$ in which both classical and quantum
       strong coupling scales behave as $|t|^{-1}$, and the quantum scale
       may be higher due to specific relationships between the parameters in
       the Lagrangian, see Ref.~ \cite{1} for an example.

\vspace{0.3 cm}

\section{Vector field model with stable NEC-violating solution.}
\label{sec:vectorfield}

\subsection{Early-time evolution: Minkowski space}

We now construct a simple vector field model which is covariant under
scaling transformation $A_\mu (x^\nu) \to \lambda^s A_\mu (\lambda x^\nu)$,
so that the Lagrangian transforms as given by \eqref{1.1}, and, furthermore,
$N\leq 4$ to avoid strong coupling. By trial and error we arrive at
the following Lagrangian
 with $N=\frac{12}{5}$ and $s=-\frac{1}{5}$:
 \begin{equation} 
     \mathscr{L}
   =q(D^{2}A^{\rho}\square A_{\rho}+kB^{2}+lC^{2}+u(\mathscr{F}_{\mu\nu}\mathscr{F}^{\;\;\;\nu}_{\rho}A^{\mu,\rho}
+2A^{\rho,\mu}A_{\rho,\nu}A_{\mu}^{\;\;,\nu}),
\label{3.3}
\end{equation}
where  $q$, $k$, $l$ and $u$ are free parameters, and
\begin{align}
\mathscr{F}_{\mu\nu} &=\partial_{\mu}A_{\nu}-\partial_{\nu}A_{\mu} \; ,
\nonumber\\
D&=A_{\mu;\nu}A^{\mu}A^{\nu} \; ,
\nonumber\\
B& =A_{\mu}A^{\nu}A^{\mu;\lambda}A_{\nu;\lambda},
\nonumber \\
C& =A^{\mu;\tau}A_{\tau}A^{\rho}A_{\mu;\rho}.
\nonumber
\end{align}
In accordance with Sec.~\ref{sec:strongcoupling}, equations of motion
have a solution
\be
A^{bg}_{\mu}=(\beta|t|^{\frac{1}{5}},0,0,0)
\label{apr14-20-2}
\ee
with constant $\beta$. This
classical evolution
occurs in a weak coupling regime at early times, $t\to -\infty$.
\vspace{0.3 cm}

We now wish
to figure out whether there exists a set of
parameters $q$, $k$, $l$, $u$ in the Lagrangian (\ref{3.3}), such that
the solution \eqref{apr14-20-2} is
stable  and violates the NEC.
By solving 
the field equation, we find 
\begin{equation*}
\beta^{5}=\frac{20u}{3m-5} \; ,
\end{equation*}
where
$$m=l+k+u.$$
To see the NEC-violation,
we need the expression for the energy-momentum tensor of this solution: 
\begin{equation*}
\begin{aligned}
T_{\mu\nu}= \frac{2\delta ( \sqrt{-g}\mathscr{L})}{\sqrt{-g}\delta g^{\mu\nu}}\big|_{g_{\rho\sigma}=\eta_{\rho\sigma}}.
\end{aligned}
\end{equation*}
To this end, we consider minimal coupling to the metric, i.e.,
set $A_{\mu; \nu} = \nabla_\nu A_\mu$,
$\square A_{\rho}=\nabla^{\mu}\nabla_{\mu}A_{\rho}$,
$D=A_{\mu;\nu}A_{\tau}A_{\lambda}g^{\mu\tau}g^{\nu\lambda},$ etc., in   
curved space-time. The Lagrangian (\ref{3.3}) can be written
in the following form:
\begin{equation*}
     \mathscr{L}
  =\frac{1}{2}f(D)\square F-f(D)A_{\tau;\sigma}A^{\tau;\sigma}+L(A_{\mu},A_{\lambda,\nu})
\end{equation*}
where
$$F=A_{\mu}A^{\mu},$$ $$f(D)=qD^{2},$$  $$L=
q[kB^{2}+lC^{2}+u(\mathscr{F}_{\mu\nu}\mathscr{F}^{\;\;\;\nu}_{\rho}A^{\mu,\rho}
+2A^{\rho,\mu}A_{\rho,\nu}A_{\mu}^{\;\;,\nu})].$$ 
We find
\begin{equation}
\begin{aligned}
&T_{00}=0,\\
&T_{ij}=p\delta_{ij}, \;\;\;\;\;\;\;  i,j=1,2,3\; , \\
&p=\Big(-\frac{1}{2}\partial_{\tau}f\partial^{\tau}F+L-fA_{\tau;\sigma}A^{\tau;\sigma}  \Big)\Big|_{g_{\mu\nu}=\eta_{\mu\nu};\;\;A_{\mu}=A_{\mu}^{bg}} \; .
\label{tij}
\end{aligned}
\end{equation}
%where $i,j=1,2,3.$
 This gives 
\begin{equation}
p=\frac{qu^{\frac{8}{5}}2^{\frac{11}{5}}5^{-\frac{12}{5}}(11-m)}{(3m-5)^{\frac{8}{5}}}(-t)^{-\frac{12}{5}},\;\;\;t<0.
\label{p}
\end{equation}
Thus, the background $A^{bg}_{\mu}$ violates the NEC provided that
\begin{equation}
q(11-m)<0 \; .
\label{4.1}
\end{equation}

Let us consider the stability of the solution $A^{bg}_{\mu}$.
Having in mind
  Ref. \cite{9}, we also require subluminality
of the perturbations  about it. Stability conditions and conditions
for the absence of superluminal perturbations for Galileon-like vector
models were derived in Ref. \cite{6}.
Making use of the results of Ref.~\cite{6}, it is straightforward to find that
there are two ranges of parameters
such that all these conditions together with (\ref{4.1}) are satisfied
for $t<0$:
\begin{equation*}
\begin{aligned}
(\RNumb{1})\;
&q>0,\\
&u\neq0,\\
&\frac{25}{2}<k\leqslant\frac{39}{2},\\
&11-k<l<-\frac{k+1}{9},\\
\end{aligned}
\end{equation*}
and
\begin{equation*}
\begin{aligned}
(\RNumb{2})\;
&q>0,\\
&u\neq0,\\
&k>\frac{39}{2},\\
&\frac{9-7k}{15}<l<-\frac{k+1}{9},\\
\end{aligned}
\end{equation*}
\vspace{0.3 cm}
Thus, our example shows that there are
stable homogeneous solutions in vector theories that
violate the NEC and avoid strong coupling regime at early times.

\subsection{Turning on gravity.}
Here we construct an initial stage of the cosmological Genesis scenario,
similiar to Ref. \cite{1}. To this end, we turn on gravity and assume
that it is described by conventional General Relativity,
while the vector field
  is minimally coupled to metric, as described above. Importantly,
all equations of motion, for both vector field and metric, remain
second order in derivatives~\cite{6}, just like in Horndeski theories.

In the asymptotic past, space-time is assumed
to be Minkowskian, and in accordance with (\ref{tij}), (\ref{p}),
energy-momentum tensor vanishes as $t\rightarrow -\infty$.
At large but finite $\vert t  \vert,$ gravitational effects on the
vector field evolution are negligible, so, to the leading order in
$M^{-1}_{Pl},$ the energy density and pressure are given
by (\ref{tij}), (\ref{p}).
Then the Hubble parameter is obtained from
\begin{equation*}
\dot H = -4\pi G(\rho+p).
\end{equation*} 
We find 
\begin{equation*}
  H=\frac{40\pi Gqu^{\frac{8}{5}}
    2^{\frac{6}{5}}5^{-\frac{12}{5}}(m-11)}{7(3m-5)^{\frac{8}{5}}}\big(-t\big)^
{-\frac{7}{5}} ,\;\;\; t\rightarrow -\infty . 
\end{equation*} 
Thus, the Universe undergoes accelerated expansion
characteristic of the early Genesis epoch. At this stage, perturbations
about the background are stable and subluminal. 
\vspace{0.3 cm}

                  \section*{Acknowledgements}

                      The
                      author is indebted to
  Yulia Ageeva, Sergey Mironov,
                      Valery Rubakov and
                      Petr Satunin for helpful discussions.
This work has been
supported by Russian Science Foundation grant 19-12-00393.

\end{document}